# The Object Coordination Class Applied to Wavepulses: Analysing Student Reasoning in Wave Physics

Michael C. Wittmann
Department of Physics & Astronomy
5709 Bennett Hall
University of Maine
Orono ME 04469-5709
wittmann@umit.maine.edu
tel: 207-581-1237
fax: 207-581-3410

## *Abstract*

Detailed investigations of student reasoning show that students approach the topic of wave physics using both event-like and object-like descriptions of wavepulses, but primarily focus on object properties in their reasoning. Student responses to interview and written questions are analysed using diSessa and Sherin's coordination class model which suggests that student use of specific reasoning resources is guided by possibly unconscious cues. Here, the term *reasoning resources* is used in a general fashion to describe any of the smaller grain size models of reasoning (p-prims, facets of knowledge, intuitive rules, etc) rather than theoretically ambiguous (mis)conceptions. Student applications of reasoning resources, including one previously undocumented, are described. Though the coordination class model is extremely helpful in organising the research data, problematic aspects of the model are also discussed.

## Introduction

For many years, researchers in physics education have focused on student understanding of specific content areas, such as mechanics, electric circuits, or heat and temperature (McDermott and Redish 1999). A common theme has been to focus more on what students are unable to do (from an expert's point of view) rather than what they are able to do and how they do it. Researchers interested in more detailed descriptions of student reasoning have slowly developed a variety of insights into student reasoning that focus on reasoning elements used by both experts and novices in varying levels of refinement (Minstrell 1992, diSessa 1993, Redish 1994). One recent proposal by diSessa and Sherin (1998), the *coordination class*, will be the focus of this article and will be described below in more detail. In broad terms, it describes one of many different possible types of concept, in which nets of simple and reasonable pieces of information are chosen and linked together.

In this article, I give a detailed snapshot of students' reasoning about wave physics in the context of a university physics class. I describe how they predominantly and inappropriately use the familiar context of reasoning about objects when thinking about many situations in wave physics. The idea of coordination classes helps me organise and reevaluate my thinking about student understanding of mechanical waves (Wittmann 1998). In previous work, I discussed how students used multiple models of wave physics to describe individual phenomena (Wittmann *et al.* 1999). I described their thinking in terms of a content-based *pattern of association* that systematised the many different reasoning elements I could see them using. In this article, I take a more general approach, focusing on more basic and less physics content-based reasoning elements in order to give a detailed account of what is meant by a conception of waves.

In the sections that follow, I discuss how students have multiple ways of thinking about and getting information about a wave system. I also describe how student responses indicate consistent use of the same reasoning resources in a variety of situations. In the first section, I summarise the elements of diSessa and Sherin's description of coordination classes. Where appropriate, I point out how their description differs from my previous idea of patterns of association. In following sections, I describe the investigations into student understanding of wave physics, how these observations show systematic similarities, and how they can be organised by defining an appropriate coordination class. In the process I raise questions about both the nature of student learning and the manner in which diSessa and Sherin's coordination classes are formed.

It is useful to apply this new theoretical description to observations of student responses because it gives a way to understand and convey the unexpected richness of student reasoning in the area of wave physics. Models used to make sense of student understanding of difficult content material must account for the seeming inconsistencies of the data. In addition, models of student reasoning should help us understand how students develop their understanding over the course of time. I hope to show that a model built up from inappropriately applied but otherwise useful and helpful reasoning resources is more productive than a theoretically ambiguous misconceptions approach in understanding our students in our classrooms.

## Theoretical Background

An ongoing debate in the field of physics education research concerns the appropriate model with which to analyse student reasoning. Although the notion of misconceptions has dominated the discussion of student reasoning in the classroom (McDermott and Redish 1999), in recent years there have been several proposals to model student thinking in physics as made up of elemental building blocks of reasoning that act as cognitive resources that are smaller and more general than misconceptions. These include phenomenological primitives (diSessa 1993), facets of knowledge (Minstrell 1992), intuitive rules (Tirosh *et al.* 1998), and reasoning resources (Hammer 2000). Whereas misconceptions are conceived as components of cognitive structure inherently inconsistent with an expert understanding, the descriptions of elemental building blocks of reasoning all have in common that they are native to both experts and novices. A particular element in these models is neither correct nor incorrect. For example, "closer means more" is correct when describing warmth near a fire, but incorrect when describing the Earth's proximity to the sun in the summer. What distinguishes novices from experts is how they apply these elemental building blocks how they activate and organise them in a particular situation. In this article, I will use the general terms *primitives* and *resources* to describe these fundamental elements of reasoning, rather than discussing the exact theoretical elements of each specific description. The two (interchangeable, for the moment) terms are meant to describe elemental building blocks of reasoning without subscribing to any specific (theoretical) description.

To illustrate in more detail what I mean both by a resource and by a coordination class made up of a set of resources, consider the following situation. You are driving and see a plastic bag in the road in front of you. Do you run over the bag, or use an evasive manoeuvre to avoid it? Your answer might depend on the ability to estimate the bag's location, size, weight, consistency, and so on. Many choices must be made to determine whether the object in the road will damage the car, and yet an expert driver is able to evaluate and act on these many different choices instantaneously. Each choice involves the activation of a resource. For example, the ability to estimate the weight of an object is one we commonly use in our interaction with the world around us, as we lift, push, lean against, or run into objects. The ability to determine location will help us determine whether or not to avoid the bag, since a moving bag is most likely empty, a stationary bag most likely full (and possibly dangerous). A novice driver will be much slower than an expert in making the appropriate decision while driving. The novice driver has the same resources as the expert driver, but the novice is unable to pull them together as quickly, and often does not have (what might be called) a compiled net of resources appropriate to this situation.

A second example can illustrate another way in which resources are chosen and combined. Consider that a car driving down the highway slows to avoid an object in its path, and is hit by another car behind it. This collision is repeated as the next car hits the second, and so on, until 20 cars have piled into each other. (Thankfully, nobody was hurt). How could one describe the pile-up? The collision can be thought of as two objects slamming into each other (or things getting in each other's way). The description of a causally linked chain reaction of events, in this case collisions, moving along a path comes naturally from our experience with doing the 'wave' in sports stadiums, yelling relay-style to pass information from one point to the next (e.g. playing telephone as children), and so on. The motion of collisions (events) through space and time is different from the motion of an object, which is physically at one location, and later at another location. It is important to note that the same resource that describes a type of interaction (collisions) plays a role in the set of resources we use to think about objects and to think about moving events.

What distinguishes experts and novices is not the existence of reasoning resources in their thinking, but the organisation of these resources into a coherent and quickly applicable whole (compare the expected behaviour of a novice and an experienced driver in the example used above). Previously, I have described the linked set of resources that students bring to a specific setting (such as driving, or the physics of waves) as a *pattern of association* (Wittmann 1998). In this linked set of reasoning resources, students apply a loose set of reasoning building blocks to describe and think about a given physical situation. Evidence of their loose linking can be found in that students often bring a set of primitives into play to answer questions on specific topics, but not always. The associations between different reasoning resources are probabilistic, not fixed, and organised in patterns researchers may be able to distinguish. diSessa and Sherin (1998) have described a similar idea as a *causal net* in which directly observable information is interpreted and leads to the inference of other information.

The interface between the outside world and what a subject observes is a filter defined by *readout strategies*. Quite literally, readout strategies define how one sees, grasps the elements, or gains information about a situation. A readout strategy describes how the observations from the outside world are parsed into meaningful elements by the subject. In the driver example above, the readout strategy might involve interpreting the size of the bag and its motion, if any (e.g. a bag blown across the pavement as cars pass it is most likely empty and therefore not dangerous), but not its colour. Together, readout strategies and causal nets describe a coordination class, in the sense that causal nets of linked resources form around the readout strategies that give the clues as to how a situation is to be interpreted.

In the following section, I consider student understanding of mechanical waves while using the theoretical description of coordination classes made up of linked primitives and reasoning resources. I describe the resources that students use, and show that students primarily use a causal net built around the readout strategy of focusing on a particle and not a series of events. The description of a common readout strategy distinguishes the coordination class model from the pattern of association described previously (Wittmann 1998).

## Investigations into student understanding of waves

In this section, I first briefly outline the research methods used to investigate student thinking about wave physics. I then describe specific findings in a set of wave physics topics: propagation, superposition, and the mathematics used to describe waves.

### *Research setting and methods*

These investigations took place while part of the Physics Education Research Group at the University of Maryland from 1996 to 1998. They were at the university level with introductory physics students in the second semester of a three-semester engineering (calculus based) sequence. These students are predominantly male (about 75%) and the majority have had high school or community college physics prior to their instruction at the university. The course includes three hours of lecture a week, a traditional laboratory, and a small group discussion section. Discussion sections consisted of either traditional TA-led recitations or tutorials, a group-learning method developed by the Physics Education Group at the University of Washington (Redish *et al.* 1997, Steinberg *et al.* 1997, McDermott *et al.* 1998). The class population consisted primarily of engineers fulfilling course requirements for their major.

The research into student understanding of wave physics took place as part of an iterative cycle of research, curriculum development, delivery, and evaluation. Among the results of the research was a set of tutorials designed to address student difficulties with the learning of wave physics (Wittmann *et al.* 2000). Research into student reasoning made use of a variety of probes, including videotaped individual demonstration interviews (with up to 30 students on each question); short, ungraded quizzes that accompany tutorials (with between 250 and 500 student responses); examination questions (with classes between 100 and 200 students); and specially designed diagnostic tests (with 137 matched students on pre and post tests). In addition, informal conversations with students during class or office hours helped form an understanding of student thought processes. Much of my understanding was built through the process of interviews. Here, individual students answer questions about a simple physical problem in a context where the researcher has the opportunity to probe their responses more deeply through extended questioning. Students who volunteer for these interviews are typically doing well in the class. Specific details of individual research investigations will be presented as needed in the discussion below.

### *Students seeing waves as objects*

The physics of waves involves the description of propagating disturbances to media, whether in finite length pulses or infinite length wavetrains (Arons 1990). In the case of mechanical waves, these disturbances can be thought of as events in which a medium is displaced from and then returns to an equilibrium state. Propagation describes the movement from one location to another of this disturbance event. Regardless of the shape of the disturbance, the physics that describes it (when using certain approximations common at the introductory level) should be the same. For mathematical convenience in physics classes, instructors and textbook authors most commonly use the idealisation of infinitely long sinusoidal waves.

However, archetypal waves of everyday experience (e.g. ocean waves) are of finite length. The results below suggest that students typically focus on interpretations based on the finite size of the waves they are familiar with. This focus manifests itself in a variety of contexts, both mathematical and qualitative, including the superposition, propagation, and reflection of waves. Student interpretations do not focus solely on the event nature of wave phenomena but on an object-like description that will be described in more detail below.

Previous investigations have looked at student reasoning in two dimensional kinematics (Grayson 1996, Snir 1989), the manner in which a wave is created and how that affects the manner of its propagation through a medium (Snir 1989, Maurines 1992), and how sound waves affect the medium through which they travel (Linder and Erickson 1989, Linder 1992, 1993). Some of the investigations described below build on previous investigations, while others investigate previously unstudied elements of how students learn wave physics.

I will begin with two contexts in which there is evidence that many students treat wave pulses as cohesive objects, rather than as extended propagating disturbances to a medium. These contexts concern questions about specific physical situations: the creation and propagation of wavepulses on a string and the superposition of wavepulses. A detailed description of a third context, sound waves, will be presented in a subsequent paper (Wittmann *et al.* 2000). In each of these situations, the types of incorrect student responses did not change. Instead, only the frequency of their occurrence was different. As a result, I do not point out when during the course of instruction the responses were given.

### Wave propagation

When students answered the question shown in figure 1 (some students answered both versions, some answered only one), a correct answer was considered to be that a decrease in the propagation time would require a higher wave speed, which could be caused by changes in the

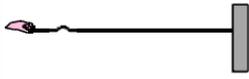

*Version 1: Free-Response (FR) format***:**  A taut string is attached to a distant wall.  A pulse moving on the string towards the wall reaches the wall in time $t_0$ (see diagram).  How would you decrease the time it takes for the pulse to reach the wall?  Explain your reasoning.

*Version 2:  Multiple-Choice, Multiple-Response (MCMR) format***:**
A taut string is attached to a distant wall.  A demonstrator moves her hand to create a pulse traveling toward the wall (see diagram).  The demonstrator wants to produce a pulse that takes a <u>longer time</u> to reach the wall.  Which of the actions *a–k* **taken by itself** will produce this result?  More than one answer may be correct.  If so, give them all.  Explain your reasoning.
a. Move her hand more quickly (but still only up and down once by the same amount).
b. Move her hand more slowly (but still only up and down once by the same amount).
c. Move her hand a larger distance but up and down in the same amount of time.
d. Move her hand a smaller distance but up and down in the same amount of time.
e. Use a heavier string of the same length, under the same tension
f. Use a lighter string of the same length, under the same tension
g. Use a string of the same density, but decrease the tension.
h. Use a string of the same density, but increase the tension.
i. Put more force into the wave.
j. Put less force into the wave.
k. None of the above answers will cause the desired effect.

Figure 1: Free response (FR) and multiple-choice, multiple-responses (MCMR) versions of the wave propagation question.  Answers *e* and *g* are correct in the MCMR question, and we considered answers like *e* and *g* to be correct on the FR question.

medium properties of the system (e.g. either an increase in the tension on the wave or a change in the mass density of the string on which the wave travelled).

Although many students at all stages of instruction include the correct response in their answers, many included responses indicating that their thinking was at odds with the usual description of waves (Wittmann 1998). These students state that the motion of the hand influences the motion of the wavepulse through the medium, e.g. that a quicker hand motion or larger amplitude hand motion will create a faster pulse. Before instruction, roughly 90% of the students gave explanations which incorrectly focused on the motion of the hand in creating the wave, and even after modified instruction, half the students still use such explanations (Wittmann *et al.* 1999).

These students seem to make an implicit analogy between the wavepulse and an object like a ball. By thinking of the whole wavepulse as an object, students can use analogies of the motion of a particle in the context of waves. One interviewed student stated the most common explanation clearly when saying 'You flick [your hand] harder...you put a greater force in your hand, so it goes faster,' and showed the hand motion he described. He moved his wrist up and down slowly to describe slow pulses and quickly to describe fast pulses. This hand motion was common to most interviewed students describing the relationship between the hand motion and the wave speed. The student's explanation was also typical of such student responses in that he was unable to give any explanation as to how the hand motion affected the speed, he merely asserted that it was the case and accepted his assertion as sufficient. Other students have given the same explanation (almost verbatim, with the same hand motions).

Students might have given the same explanation to describe how to throw a baseball so that it travels more quickly (or more slowly). Their description focused more on the 'tossing' of the wave into the system than on the movement of the wave through the system. This explanation is consistent with a description of waves as objects rather than a description of waves as a propagating disturbance within a system. Similar explanations were given on the many written tests administered both before and after instruction on waves.

## Wave superposition

In the question shown in figure 2, a correct answer was expected to show that the wavepulses pass through each other. Superposition of the waves during part of their motion should have no permanent effect on the wavepulses (in the commonly used small angle approximation). The two most common responses to the question are shown in figure 3. A notable minority of students gave the incorrect answer shown in the figure. As one student wrote, 'part of the greater wave is cancelled by the smaller one'. In explanations, students wrote as if they were describing a collision between objects. For example, if two unequal size and mass

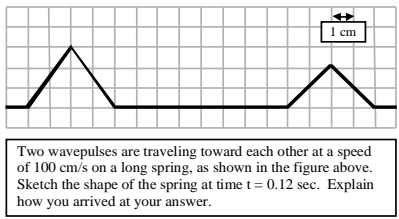

Two wavepulses are traveling toward each other at a speed of 100 cm/s on a long spring, as shown in the figure above. Sketch the shape of the spring at time t = 0.12 sec. Explain how you arrived at your answer.

Figure 2: Wave superposition question in which students are asked for the string shape at a time after the waves have overlapped.

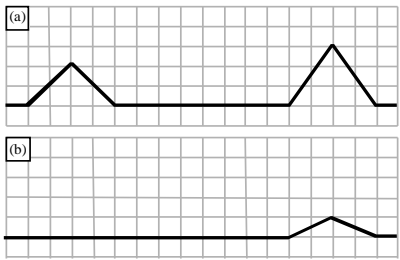

Figure 3: Common responses to question in Figure 2. (a) Correct response showing the wavepulses having passed through each other unchanged. (b) Common response showing cancellation of wave amplitude.

carts moved toward each other at the same speed and collided in a perfectly inelastic collision, then the unit of two carts would move in the same direction as the more massive of the carts. The speed would be much slower. One student's written comment (on a similar question asked during a later semester) supports this interpretation, 'the smaller wave would move to the right, but at a slower speed'. In other questions, students sometimes described identically shaped, symmetric wavepulses bouncing off each other. This is consistent with the visual interpretation of observations of two identical wavepulses passing through each other as they move in opposite directions along a string. In both these questions, students describe wavepulses as if they were localised objects, and do not focus on the events surrounding their propagation or interaction.

### Students seeing wave peaks

In the following examples, I focus on student use of single points to guide their reasoning about a wavepulse as a whole. These examples, from the mathematics used to describe waves travelling through systems and superposition of waves, both demand an understanding of waves as extended regions of disturbance. Many students instead describe waves as if only one point of the wave is important. The ability to use a single point to describe a larger object is often important, and students at this stage of physics instruction have had much practice applying it to a variety of situations (for example, in kinematics or dynamics, where one focuses on the centre of mass of an object).

## Mathematical description of propagating waves

The question in figure 4, which I will refer to as the wave-math problem, asks students to interpret a physical situation represented by a mathematical formula. A correct answer (see

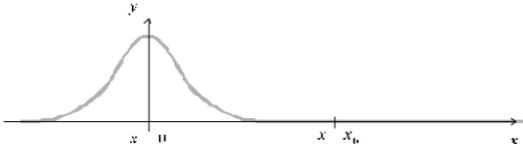

Consider a pulse propagating along a long taut string in the $+x$-direction. The diagram below shows the shape of the pulse at $t = 0$ sec. Suppose the displacement of the string at this time at various values of $x$ is given by

$$y(x) = Ae^{-\left(x/b\right)^2}$$

A. On the diagram above, sketch the shape of the string after it has traveled a distance $x_0$, where $x_0$ is shown in the figure. Explain why you sketched the shape as you did.

B. For the instant of time that you have sketched, find the displacement of the string as a function of $x$. Explain how you determined your answer.

Figure 4: Wave-math question in which students must correctly interpret both physical and mathematical representations of propagating waves

figure 5-a) would state that the shape of the wavepulse was unchanged, but its centre was now located at $x_0$. Mathematically, this would be described by rewriting the equation with $x$ replaced by $x-vt$, where $x$ describes the location of a piece of string, $v$ is the propagation velocity of the disturbance in the system, and $t$ is the time after an arbitrarily chosen starting point (in this case, the moment when the amplitude point passes the origin of an arbitrarily chosen coordinate system). At a later time, $t$, higher $x$ values give the same value for $x-vt$ as at $t = 0$ in the original equation.

A common answer was to draw a wavepulse centred at $x_0$ but with a lower amplitude (see figure 5-b). Most students who sketched a pulse whose amplitude had decreased gave the explanation that the exponent value would decrease as the value of $x$ increased. One student first stated,

> Okay. Umm. Let's see. Sketch the shape of the spring after the pulse has travelled [Mumbling as he re-reads the problem]. Okay. Over a long, taut spring, the friction or the loss of energy should not be significant; so the wave should be pretty much the exact same height, distance, -- everything. So, it should be about the same wave ... it's got the same height, just a different $x$ value.

This excellent explanation suggests a clear understanding of the physics and the various issues that might play a role in the physics of the situation. For example, the student states that losses due to energy (which might contribute to a lower amplitude in a real-life situation) do not play an appreciable role to the idealised situation in this question. Note, though, the use of the term 'different $x$ value,' by which the student seems to imply that the location of the peak is at a different $x$. The student then continued,

> No, wait. ... It doesn't say that $y$ varies with time, but it does say it varies with $x$. So that was my first intuition but then, looking at the function of $y$. Let's see, … I guess it'll be a lot smaller than the wave I drew because the first time x is zero, which means A must be equal to whatever that value is, because e raised to the zero's going to be 1. So, that's what A is equal to. And then as x increases, this value, e raised to the negative, is going to get bigger as we go up. So, kind of depending on what v is. Okay. So, if x keeps on getting bigger, e raised to the negative of that is going to keep on getting smaller ... So the actual function's going to be a lot smaller.

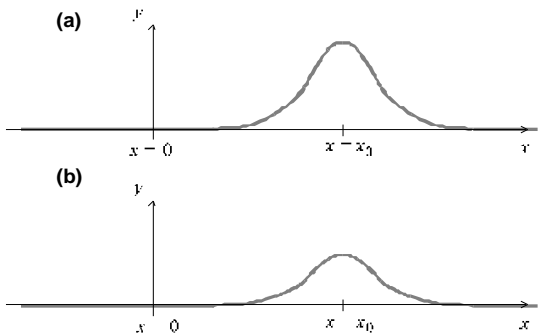

Figure 5: Correct responses to the wave-math problem in Figure 4.
(a) A correct sketch of the shape of the pulse at a later time,
showing the amplitude unchanged. (b) Common sketch of the shape
of the pulse showing the amplitude decreased - typically
accompanied by reasoning which misinterprets the mathematics.

The student is misinterpreting the physics equation and in the process reevaluating and changing a previously correct response. This response was typical of (though far clearer than) those given by students who drew the wavepulse with a lower amplitude.

Setting aside for a moment a discussion of the student seeking consistency between the mathematics and classroom observations of similar experiments, consider the role the mathematical equation played in altering the student's response. Students typically cite the equation and the effect of a change in $x$ on the exponent. In the quote above, the student confuses $y$ with $A$, the amplitude. In addition, there is evidence of a particular type of reasoning resource, which Sherin (1996) calls 'symbolic forms'. Here, students apply the form $e$-(exponential decay) to the *amplitude* of the wave function, as a decay in its peak value over time, rather than to the *shape* of the function and its decay in value at one time. Their focus, it seems, is on the peak of the wavepulse, rather than on the whole of the wave.

## Wave superposition

Students also often use only a single point on the wave when describing the physics of wave superposition. The most common answers to the question shown in figure 6 are shown in figure 7. A correct answer (figure 7-a) would show point-by-point addition of displacement at every location where the individual wavepulses overlap.

Students giving the responses shown in figure 7-b and 7-c seem to be focusing inappropriately on only a single point of the wave in their descriptions. A student who drew a sketch like the one in figure 7-b explained, 'The waves only add when the amplitudes meet'. Unless the two points of the wavepulses that the student considers relevant overlap, these students assume there is no summation of displacements (superposition) in the region where the wavepulses do overlap. The extended region of the wave was described by a single point, and no other point mattered.

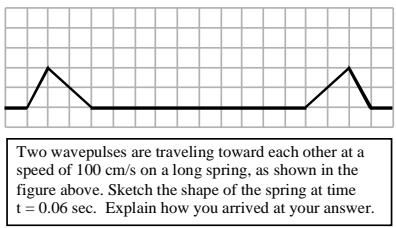

Two wavepulses are traveling toward each other at a speed of 100 cm/s on a long spring, as shown in the figure above. Sketch the shape of the spring at time t = 0.06 sec. Explain how you arrived at your answer.

Figure 6: Wave superposition question in which students are asked for the string shape at a time when the peaks do not overlap.

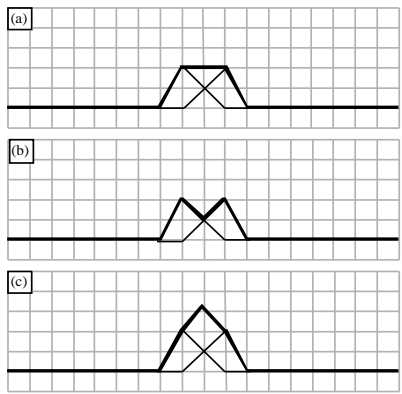

Figure 7: Common responses to question in Figure 6. (a) Correct response showing point-by-point addition of displacement at all points. (b) Common response showing no addition when peaks do not overlap, (c) Common response showing addition of peaks even though they do not overlap.

A different answer given for similar reasons is shown in figure 7-c. A student who drew a sketch like this explained, 'since the waves are on top of each other, the amplitudes add'. This student appears to be describing the interaction of two extended regions of displacement by focusing only on the two amplitude peak points. Addition was not explained nor considered for any other points of the wavepulses (i.e. the sketch indicated addition away from the peak points but was indicative of the peak points adding).

Both types of responses were repeated nearly verbatim by many students in interviews. Many students often were unable to explain through more than an assertion. It seemed that the assertion itself was sufficient as an explanation for these students. At no time did these students discuss the general principle of superposition, where the displacement at each location due to each wavepulse would be added. Still, when the peaks overlapped, they did show addition of the peaks. Their understanding and use of superposition seems confined to the peak point alone.

## The Application of the Object Coordination Class to Wavepulses

The investigations described above and others in the literature have shown that wave physics is difficult for many students to learn. To describe their reasoning only in terms of difficulties, though, ignores the richness of thinking that the students are using to produce their answers. To show the productive reasoning students employ, I will describe both the causal net and the readout strategies that seem to play a role in leading students to this reasoning. Students use many resources at the same time, consistently bringing multiple primitives to a single wave physics topic. As illustrated in the examples above, the typical student causal net revolves around a readout strategy that focuses on an interpretation of wavepulses in terms of a single object. This object may be described in terms of a single point (the peak point of the pulse). At the same time, students bring resources dealing with events and obvious wave properties to their thinking. The manner in which students use contradictory resources will be discussed below.

### *A causal net of student reasoning resources*

The reasoning elements students use in wave physics can be described in terms of the phenomenological primitives (diSessa 1993) and facets (Minstrell 1992) that have been described in the literature. Other, easily described resources also play a role in student reasoning.

#### Creation and motion primitives

A variety of primitives describe the creation and motion of wavepulses. Based on their description of creating faster or slower wavepulses, students seem to be thinking of an Actuating Agency (Hammer 1996) to describe the force needed to set an object in motion. Without the hand motion, no wave would exist. Once the wave is created, no Maintaining Agency (Hammer 1996) is needed to keep it going, it simply moves through the system. Both these descriptions are relatively consistent with the description of a ball being thrown through the air (until it falls to the ground).

The effort exerted determines the speed of the moving object (the wavepulse) for many students, which is consistent with Working Harder (diSessa 1993). When additional force is exerted on a wavepulse, it moves faster; when less force is exerted, the wavepulse moves more slowly. Often, these students state that the faster wavepulse is also larger, due to the additional force exerted in its creation.

Conversely, some students using changes in hand motion to describe how to change wave speed state that smaller wavepulses will move more quickly through the medium. This is analogous to the belief that a small mouse will move faster than a large elephant (which is true, compared to body size, but false compared to actual distance covered in the same amount of time). This primitive has been described as Smaller is Faster by diSessa (1993).

Some of these resources could play a role in thinking of waves as propagating events. For example, Actuating Agency correctly describes one manner of creating a wave. Working Harder gives the proper description of the amplitude of a wave (in the wave propagation example), while not affecting the wave speed (though giving the illusion of doing so). Smaller is Faster reasonably describes fast waves created on especially taut strings, where it is difficult to create larger amplitude or larger width waves. Note, though, that students are not using the resources in an appropriate fashion for event-like descriptions of waves. They are focusing on object-like interpretations of these responses.

#### Interaction and time-evolution resources

Other reasoning primitives are also observed in students' descriptions of wave physics. For example, Bouncing describes one type of collision, as objects bounce off each other because they get in each other's way. The Bouncing primitive (diSessa 1993) manifests itself in a variety of fashions, such as 'equal size waves bounce off each other' while the similar Cancelling primitive manifests itself when students say 'smaller waves cancel out parts of larger ones'. One complicated refinement (an example of coordination of different resources) comes when students describe the resulting smaller wavepulse as moving slower. In this case, the description is inconsistent with the previously described use of the Smaller is Faster primitive by some students.

Another interaction resource, only loosely defined in this context, is Addition (or Combining). Students use this resource in a variety of ways when describing overlapping wavepulses, either to describe overlapping bases, peaks, or individual points of the wavepulses. I will not define this resource in more detail, since the reader can easily imagine countless examples of its use in mathematics and other settings.

A further mathematical resource is the mathematical form described earlier (exponential decay). This resource is consistent with the simplification of Dying Away (diSessa 1993). This describes the common phenomena that all moving objects eventually stop moving. Similarly, a propagating wavepulse eventually dies away (in the real world, due to dispersion in the system and friction both internally and with the surrounding world).

Again, these resources make sense in a wave or event-like context. Waves often seem to bounce off each other (though they do not cancel). Waves also add together. And, waves do die away, finally, due to friction and energy loss within the system. The use of these resources alone is not problematic.

### Readout strategies to guide student reasoning

Since the individual resources students bring to the description of waves are not unique to the object-like description of waves we observe them using, a different element of their reasoning must be considered. The manner in which they choose to use resources must be evaluated.

## A guiding readout strategy: Object as Point

Earlier, I described readout strategies as the manner in which people literally see the system. Fundamentally, many students' readout strategies seem centred around choosing a point interpretation of the wavepulse. This interpretation both influences and is influenced by the resources that students have at their disposal. The peak of the pulse is an obvious visual cue; it seems natural to focus on it when describing mathematics or wave interactions. The mathematical description of waves requires that something decay with position; it seems natural that the amplitude of the wave fill that role. The superposition of wavepulses requires some sort of addition; it seems natural that it only occur at (or with) the peak points. The creation of a wavepulse requires some sort of exertion; it seems natural that additional exertion should create faster waves just like when throwing a ball (which we can think of as a point particle).

Another way to state this readout strategy would be, 'I can think of an extended object by focusing on only one point'. This simplification helps when analysing a large set of introductory physics problems, such as those dealing with the trajectory of most real-life objects. In such situations, treating the object (a person shot from a circus cannon, for example) as a point allows general and simple equations of motion to be applied. In solving such a problem, one can ignore the rotation of the person about their centre of mass, the flailing of arms and legs, and so on. Similarly, a common skill taught to students at the introductory level is to describe an object as a single point when sketching a free body diagram. Thus, an object's location and its interactions with its surroundings are often described by focusing on a single point.

Outside of physics, the same resource is often used without question. When describing football players as points on the screen during televised football games and aeroplanes as dots on a radar screen, the representation is natural and sensible. A further example is the literary term *synecdoche*, in which a part or characteristic of an object or phenomenon represents the whole. An example is the sentence, 'A million faces were turned to the TV at that moment'. In this sentence, each person's face is used to represent each person and that person's attention being focused on the events on the TV. Similarly, a president represents a country, a CEO the company, and so on.

In the context of wave physics, the Object as Point readout strategy is a natural way to describe waves. Though a wavepulse is an extended region of displacement from equilibrium, the peak naturally serves to describe its location. In both the wave-math problem and the superposition questions, students focused on a single point rather than describing each individual piece of the system. They applied the point readout strategy inappropriately, employed resources in an object-like fashion, and thereby missed aspects of the wave nature of the system.

## Additional readout strategies

In addition to a readout strategy that focuses only on the peak of the wavepulse, we observe students using other properties to decide that wavepulses are object-like. The creation of a wavepulse requires an action like a ball throw; it seems natural that the subsequent motion of the wavepulse would be dependent on the manner of the 'toss'. The interaction of wavepulses often seems to consist of the pulses bouncing off each other; it seems natural to assume that this is the case if no other cues exist for interpreting the information more accurately.

Students may approach the infinite length (sinusoidal) wavetrains that they most regularly encounter in the physics classroom using similar reasoning. One student reasonably described sinusoidal waves as a succession of wavepulses. The inappropriate readout of sinusoidal waves as a succession of object-like wavepulses may not be resolved for many students for a variety of reasons. Many times, students are asked only to find the superposition of sine waves when the peak points perfectly overlap. Students do not have to interpret the displacement of the wave at all points, but only at the amplitude (and zero displacement) points. In certain wave physics contexts, students described sinusoidal sound waves exerting forces on the medium through which they travel. This gives a hint of the object coordination class being applied to infinite length waves, but at the moment, the data are unclear and incomplete.

## Evidence of student coordination

Students in this study used multiple resources and readout strategies in their reasoning. To illustrate the organisation of such resources, figure 8 shows the first linking layer of a graphical representation of the causal net of student object-like descriptions of waves. Several cues, when combined, may bring up several resources all at once. Each individual resource makes sense within an individual context, so the set of resources linked together might seem reasonable, too. The linked resources can be used without much thought and without much further interpretation. Thus, the movement, interaction, and mathematical interpretation of these wave objects is simplified for students.

In one interview in which students answered a large number of questions on wave physics (including those described in this article), a student (David) explicitly used most of the above-described reasoning resources in an object-like fashion.

For example, when answering the wave propagation question (see figure 1), he stated, 'I think possibly, you see a slower … pulse if the force applied to the string is reduced, that is: the time through which the hand moves up and down [is reduced]'. David's response is consistent with Working Harder for an object that moves slower due to a lesser force. The action of wavepulse creation affected his response for wavepulse propagation.

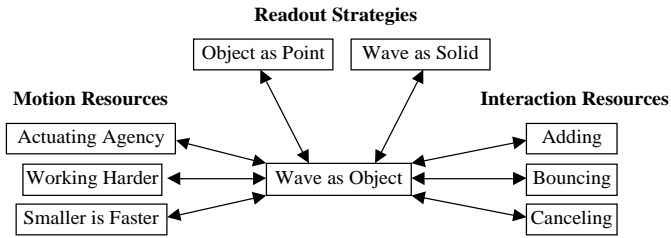

Figure 8: Possible schematic showing reasoning resources that describe an object-like model of waves. The interplay of different elements (such as Wave as Solid with Bouncing) might explain how students answer certain wave physics questions.

In the context of the wave-math problem, he described a decaying propagating wavepulse. He stated that the shape would stay the same, but the amplitude would change, and said 'this is a decay function' while pointing to the equation given in the problem. This shows clear evidence of the exponential decay form (in this case, an interpretation of the dying away primitive in a mathematical context). In addition, his response shows evidence of the Object as Point readout strategy playing a role in his interpretation of the physics.

The set of primitives that David brings into play in wave physics require that he think of waves as objects at times, and that these objects can be described by a single point when needed. In a sense, he is describing a new type of object, one that is often like others treated in physics (i.e. moves, bounces, its motion dies away, conveniently described by a single point).

But, David also shows that this object is not quite like other objects. David explicitly and clearly used both object properties and superposition (i.e. the addition of displacement) when he described the interaction of two wavepulses. In his first response to the superposition question in figure 6, he stated that the wavepulses would bounce off each other. This clearly indicates that he was thinking of wavepulses as objects that bounce, in essence like solid billiard balls. When asked to explain his answer, he changed his mind and then sketched a response like the one shown in figure 7-b. Here, he was focusing only on the point of the wavepulses. Only when asked to describe overlapping amplitude peak points did he describe superposition correctly, and then only for a limited case. Where an expert physicist would think of these problems in only one way, David had three separate ways of thinking about the physics.

David also answered many questions correctly when describing wave physics. Thus, he used the same reasoning resources in a variety of settings, sometimes object-like and sometimes event-like. In more advanced settings, using specially designed diagnostic test questions, we have found that students consistently use multiple types of explanations in their responses on a single test. This use of resources in multiple fashions will be discussed in more detail below.

## Student Reasoning Within a Coordinate Class

David's use of non-object properties while giving primarily object-like responses raises questions about the nature of coordinate classes. Superposition (addition, when applied to the shape of the wave) is a property that simply does not exist for objects (outside of science fiction movies). Two pieces of chalk are never 'on top' of each other, they never 'add' at a given location in space.

This added property in an otherwise object-like wave raises certain questions. These include: What sort of object are students describing, if it is capable of superposition? How does a property fundamentally inconsistent with a class come to be added to that class? Shouldn't the students realise that they're being inconsistent and experience some sort of cognitive conflict in their thinking?

These questions have a set of possible solutions that could be proposed, though I speculate in each case without sufficient evidence to come to a conclusion. The suggestions are therefore described both to challenge readers interested in using coordinate classes to describe student reasoning and to suggest the need for further refinement of studies into student thinking. I will make two suggestions that might help solve the issues raised by the existence of inconsistent properties within an otherwise productive class of reasoning. The first is that coordinate classes are not always firmly in place in our reasoning, and that a 'just-in-time compile' of a variety of reasoning resources occurs as (specifically) students answer questions with which they are unfamiliar. The second is that students lock into a certain type of class based on their (possibly unconscious) readout strategies and are more inclined to make perturbations to this class than switch classes wholesale.

### The just-in-time proposal

When students encounter wave physics questions, they have most often just completed at least a semester's worth of instruction on mechanics. In mechanics, most reasoning (at the introductory level) involves only objects, and most objects are simplified to a single point in most problems. Thus, students coming to the study of waves are primed to reason about objects in physics. That they do so in wave physics should not come as a surprise.

Thus, when students encounter additional properties, such as superposition, they do not question the object basis of their reasoning. Instead, they add properties to what seems otherwise reasonable. These properties are added 'on the fly,' without much consideration. After all, focusing on only the single point that indicates the peak of a wavepulse (or a sinusoidal wave) is enough when trying to understand typical diagrams shown in textbooks where superposition is shown only for perfectly overlapping waves, causing either perfect constructive or destructive interference.

Support for the just-in-time description of student reasoning comes from the ease with which students bring up and toss aside primitives while thinking. David, above, gives one description of interacting waves, based on collisions, then jumps to another (only amplitude point addition). The student in the wave-math study first used excellent physics intuition, then suddenly applied the exponential decay form to reinterpret everything just stated. In the wave propagation questions, students often first apply object-like resources related to the hand movement, then give other wave physics explanations (based on tension and string mass density) (Wittmann *et al.* 1999). Students are not only capable of holding multiple interpretations of single physics situations, they are also capable of switching between different interpretations at a high speed, while not judging the inconsistencies between their answers. Understanding the timescale of student reasoning might play a fundamental role in answering these questions (von Aufschnaiter 1999).

This use of 'on the fly' reasoning, including new properties only when essential, otherwise relying on typical modes of thinking, raises questions about the nature of coordination classes.

How robust are they? How are they created? When do students' coordination classes shift from modifications of an old class to the creation of whole new classes? What is the role of student epistemological knowledge within the coordination classes?

In addition, there are certain implications for instruction. For example, students may not be aware that they are using incompatible reasoning. The role of an instructor might be to create a classroom environment in which the consistency of reasoning would be emphasised. Such a classroom would focus on students making their thinking explicit and comparing the different elements of their reasoning. This would require instructors to help students develop skills which may not be taught in the traditional (nor in our modified tutorial) classroom. If the just-in-time compile assumption is correct, students in a modified 'consistency checking' classroom should perform better on questions such as the ones discussed in this article than did the students in the studies which form this article. As such, the effect of just-in-time thinking on the student's part could be experimentally measured.

### The inability-to-shift-classes proposal

Another way to describe student reasoning when discussing the object coordination class in the context of wave physics is to compare the object coordination class to the one we wish they would use. It may be that the students are completely unaware of the choices that they are making as they describe waves in object-like terms, and that this lack of awareness plays a profound role in their ability to think about the physics.

Rather than thinking of waves in terms of wavepulses, students should focus on the event properties of a propagating disturbance to a system. In addition to the event-like situations described above, there are others that students are familiar using: news travelling (motion of information without it leaving its original source), sports crowds cheering and then falling silent (fading in, fading out), and so on. The focus on different fundamental classes of reasoning resources might lead one to studies such as the ones done by Michelene Chi and colleagues (Ferrari and Chi 1998) on the classes of ontologies which form the basis of our reasoning. Another area in which studies might be profitable would be the study of theory change, based on Kuhn's work on scientific revolutions but applied to individual students and their use of theories in their reasoning (Strike and Posner 1992, Wellman 1998).

The difference between object and event might be important for students to see in the classroom, simply because both of these classes are so fundamental that a traditional elicit-confront-resolve approach might not work. It is simply impossible for students to stop thinking of objects; instead they would have to refine their thinking such that they realised that waves fell into a wholly separate class of reasoning than the one they were originally bringing to the classroom.

Again, there would be implications for instruction if this proposed situation existed. For example, we could create a situation in which explicit cognitive conflict were created in students such that they realised that the object coordination class was not the most effective at describing wave physics most generally. As noted above, the object coordination class often seems to be used exactly because it is sometimes useful. Instruction would have to focus on those areas where an event description would be more appropriate. Thus, students could be helped to develop an understanding of waves based on an event coordination class (however this may be defined).

It may be, though, that the shift from object to event may not depend on cognitive conflict caused by confrontation. A refinement of object reasoning may be more appropriate: rather than focusing on the *wave* as object, students might focus on the objects that make up the system. Thus, the motion of the hand (an *event*) might be related to the motion of the first little piece of string, which might be related to a later piece of string, and so on. This might provide a bridge for students to think about the interactions of elements of the system which lead to propagation. Again, as in the case of the just-in-time compile, the different instructional methods should have experimentally measurable results.

## Discussion

When studying student understanding of a physics content area, an individual student response is often interpreted in terms of a single model or misconception. Students are assumed to have only a single model of the physics. Results described above show that such a description is not necessarily productive, nor is it complete. Instead, the results show that students use reasoning elements both inappropriately and appropriately at the same time. To describe this as a single robust model misrepresents the data.

The description of an incorrect student approach to wave physics in terms of the inappropriate use of the object coordination class has several implications for both researchers and instructors in physics.

Researchers must be aware of the elements of student reasoning as they draw conclusions from their data. Students have many ways of building their understanding at the moment that it is needed. For example, the object coordination class applies to student reasoning about wavepulses, but they may reason very differently about wavetrains. Researchers cannot assume that the two form a unified model in the students' heads, nor that students will consistently think about wavepulses alone. Instead, researchers might see whether students have multiple views of specific content areas and find ways to describe this. Also, researchers could focus on developmental issues, how students build causal nets, why they focus on certain elements of systems, and why they believe the answers they give. A great variety of questions remain. Rather than restricting possibilities for research, a description of student reasoning in terms of coordination classes allows more refined and exact questions to be asked, even as issues about coordination classes themselves are clarified.

The fleeting nature in which some primitives are used indicates that students do not have a robust model that they are applying to the wave physics. Instead, they seem to construct their understanding on the spot, possibly a type of just-in-time building of what seems sensible and helps them answer the question. The timescales of student reasoning are presently unclear.

Another interpretation to these data would be that students are applying the object coordination class to waves, but are unsure of their responses and are thus willing to change their answers quickly and often in order to find the one that feels most comfortable to them. What consistency checks are being used in such a situation? diSessa and Sherin have not adequately described the role of epistemologies in determining student readout strategies, the choice of coordinate classes, and the decision to switch between classes. The data in this article also do not give any insight into the question. Further research would be required to adequately understand the relationship between student epistemologies and coordination classes.

Similarly for instructors, the focus on student coordination classes allows more detailed questions about the instructional methods that might be appropriate for a certain subset of students. As Hammer has pointed out (1996), the needs of individual students and the goals of the instructor call for different teaching methods to exist in flexible, malleable, and changeable form in the instructor. Providing the correct learning environment for a student involves recognising the complexity of student reasoning. One model of teaching might be to consider student learning in terms of constructing the appropriate coordination classes. Part of this task would be helping students evaluate the usefulness of the primitives they bring to the classroom. This might involve helping students deconstruct an inappropriate causal net, leading students to an understanding of the appropriateness of individual reasoning resources to a content area, or guiding students to the use of primitives which can be linked in a useful fashion (Clement 1982). Another model of teaching might involve helping students determine which of their coordination classes are most appropriate in a given situation. The refining of everyday thinking (using events, not objects, to think about waves, and understanding why that choice is made) might then be an explicit goal of the instruction (Arons 1990). Determining the best of these methods or the appropriate mixture of them would be an area for action research within an individual classroom.

## Conclusion

In this article, I have tried to show that one can describe student reasoning in wave physics in terms of reasoning resources that are inappropriately linked together into an object-like model. The use of the object coordination class to describe wavepulses is typified by an inappropriate interpretation of the wavepulse as a single, unified, pseudo-solid object and the description of this object by a single point.

The object coordination class (and the point primitive to describe the most important part of that object) when applied to wave physics allows correct reasoning in some cases (for example, the superposition of sinusoidal waves when their amplitude points perfectly overlap) but fundamentally misinterprets the wave physics. In addition, students add properties wholly inconsistent with objects in order to describe additional, unfamiliar elements of wave physics. The manner in which they do so raises questions about the nature of coordination classes, the manner in which students make (possibly epistemological) choices in their reasoning, and the manner in which transitions are made between classes. These issues need to be addressed in future research.

In this article, I have not discussed curriculum materials that could help students come to a better understanding of waves (Wittmann *et al.* 2000). I have also not showed how students develop the wavepulse coordination class (though I have suggested the possibility of an on-the-fly or just-in-time construction based on what seems reasonable to the student at that instant). I suggest that further research with more detailed data is needed to gain more insight into the finer points of this extremely productive theoretical model of student understanding.

## Acknowledgements

This work was done while a member of the Physics Education Research Group at the University of Maryland, and I am indebted to all its members for contributions to this work. In particular, I would like to thank David Hammer and Andy Elby for their detailed comments and discussions and the three reviewers whose comments have helped make this a much better article. Richard Steinberg and E. F. 'Joe' Redish provided guidance and mentorship in developing the ideas that play a role in this article. Lei Bao, Mel Sabella, and Jeff Saul helped collect specific data reported here. This research was funded in part by National Science Foundation grants DUE-9455561 and DUE-9652877 and Department of Education FIPSE grant 116B70186.